# Energy exchange and localization in essentially nonlinear oscillatory systems: canonical formalism.

O.V.Gendelman[1], T.P.Sapsis[2]

1- Faculty of Mechanical Engineering, Technion – Israel Institute of Technology

2 – Department of Mechanical Engineering - Massachusetts Institute of Technology

Over recent years, a lot of progress has been achieved in understanding of the relationship between localization and transport of energy in essentially nonlinear oscillatory systems. In this paper we are going to demonstrate that the structure of the resonance manifold can be conveniently described in terms of canonical action-angle variables. Such formalism has important theoretical advantages: all resonance manifolds may be described at the same level of complexity, appearance of additional conservation laws on these manifolds is easily proven both in autonomous and non-autonomous settings. The harmonic balance - based complexification approach, used in many previous studies on the subject, is shown to be a particular case of the canonical formalism. Moreover, application of the canonic averaging allows treatment of much broader variety of dynamical models. As an example, energy exchanges in systems of coupled trigonometrical and vibro-impact oscillators are considered.

## 1. Introduction

Canonical action – angle (AA) variables are famous and widely used instrument in a theory of dynamical systems [1 - 4]. The AA variables were instrumental in formulation of many prominent results and theories. Among others, one can mention theory of adiabatic invariants [1], formulation and proof of KAM theorem [3, 5, 6], development of canonical perturbation theory [5, 6], explorations on Hamiltonian chaos [7, 8], autoresonant phenomena [9, 10] etc.

The issue of energy exchange and transport in oscillatory systems recently attracted a lot of attention. Among various physical problems, considered in this context, one finds targeted energy transfer in essentially nonlinear systems [11-14], wave propagation and energy transport in granular media [15, 16], discrete breathers in strongly nonlinear systems, as well as vibration absorption and mitigation provided by nonlinear energy sinks [17, 18]. Major progress in all these fields has been achieved, since it was realized that the most efficient energy transport in the oscillatory systems usually occurs in conditions of resonance. This observation allows one to treat the system in the vicinity of the resonance manifold, and to restrict the consideration by averaged equations of motion (usually referred to as slow-flow equations). This crucial simplification often allows reduction of dimensionality and gives rise to conservation laws absent in the complete system beyond the resonance manifold. Technically, in vast majority of the mentioned works, the averaging has been performed with the help of complex variables (complexification-averaging approach, CxA) [19 - 21]. This approach follows back to models with self-trapping [22] and rotating-wave approximation [23] in the lattice dynamics. From mathematical point of view, this approach is completely equivalent to classical harmonic balance with slowly varying amplitudes [24]. However, the formalism of CxA allows convenient handling of the slow-flow equations. Advantages of this method were demonstrated in recent works devoted to energy exchange in model oscillatory systems [25, 26].

The goal of the current work is to present the formalism based on the canonical AA variables that allows efficient treatment of the energy transfer problems. Moreover, we are going to demonstrate that the CxA formalism is a particular case of this canonical AA formalism. Strictly speaking, the complex variables used in CxA naturally arise from transition to the AA variables of the linear oscillator. Exploration of the dynamics on the resonance manifold in terms of the AA variables easily reveals all regularities mentioned above (reduction of the state space, additional conservation laws). Besides, in terms of the AA variables one can study the energy transport in systems not amenable for the CxA (or harmonic balance) treatment, such as vibro-impact oscillatory systems.

## 2. *Low-dimensional systems.*

In order to present the approach, we first consider the simplest possible nontrivial settings for the energy exchange in oscillatory systems: conservative system with two degrees of freedom and a single conservative nonlinear oscillator with external forcing.

*2.1 Conservative system with two degrees of freedom.*

Let us consider the conservative system of two coupled oscillators. Generally speaking, the Hamiltonian of this system is expressed as:

$$H = H(p_1, p_2, q_1, q_2) \tag{1}$$

Here $q_k, k=1,2$ are generalized coordinates, and $p_k, k=1,2$ are conjugate momenta. It is supposed that at given energy level System (1) occupies a finite hypersurface fragment in the state space.

We'll say that the canonical transformations to action-angle variables are *induced* by single – DOF Hamiltonians $H_0(p,q)$ with sets of periodic solutions parametrized through their energy levels $H_0(p,q) = E = \text{const}$ [1, 4]. Then, the action-angle variables are defined by the well-known formulas [4]:

$$I(E) = \frac{1}{2\pi} \oint p(q,E)dq; \quad \theta = \frac{\partial}{\partial I} \int_0^q p(q,I)dq \tag{2}$$

By inverting expressions (2), one can get explicit formulas for the canonical change of variables $p(I,\theta), q(I,\theta)$. Each particular Hamiltonian $H_0(p,q)$ generically induces canonical transformation of this type. For each conjugate pair of variables in Hamiltonian (1) we use one of such transformations:

$$\begin{aligned} &p_k = p_k(I_k, \theta_k), \; q_k = q_k(I_k, \theta_k), \\ &k = 1,2; \; I_k \in [0,\infty), \; \theta_k \in [0, 2\pi) \end{aligned} \tag{3}$$

It is not required that the transformations for different *k* will be induced by the same Hamiltonian and will have the same functional form. As a result of the transformation, the system will be described by the following Hamiltonian in terms of the action – angle variables:

$$H = H(I_1, I_2, \theta_1, \theta_2) \tag{4}$$

Due to a $2\pi$-periodicity of the angle variables, it is possible to expand the Hamiltonian into Fourier series [7]:

$$H(I_1, I_2, \theta_1, \theta_2) = \sum_{m,n} V_{m,n}(I_1, I_2) \exp(i(m\theta_1 - n\theta_2)), V_{m,n} = V^*_{-m,-n} \tag{5}$$

Averaging procedures in Hamiltonians similar to (4) are always based on existence of slowly varying combination of the angle variables. Commonly, this slow phase exists due to the fact that the actions do not deflect much from their average values [7]. It will be demonstrated below that the slow phase may appear also due to other reasons. At this stage, we proceed formally and suppose that the phase variables combine into a single slow phase $\vartheta = m_0 \theta_1 - n_0 \theta_2, m_0, n_0 \in \mathbb{Z}$. Averaging the Hamiltonian over the fast phases, one just removes from (5) all terms not proportional to the slow phase, substitutes the actions by their average values and then obtains a slow-flow Hamiltonian in the following form:

$$\bar{H}(J_1, J_2, \vartheta) = \sum_p V_{m_0 p, n_0 p}(J_1, J_2) \exp(ip(m_0 \theta_1 - n_o \theta_2)) = \sum_p V_{m_0 p, n_0 p}(J_1, J_2) \exp(ip\vartheta)$$
$$J_k = \langle I_k \rangle, \ k = 1, 2 \tag{6}$$

Formally, the summation in (6) should extend over all integers. In practically interesting cases, due to fast decrease of the Fourier coefficients, it might be sufficient to consider only small values of *p*. We take some freedom in calling expression (6) "Hamiltonian", since the slow variables do not form the canonically conjugate pairs. Precisely speaking, the function introduced in (6) is an integral of motion for the slow variables, rather than the Hamiltonian. Evolution equations for these variables will take the following form:

$$\dot{J}_1 = \left\langle -\frac{\partial H}{\partial \theta_1} \right\rangle = -m_0 \frac{\partial \bar{H}}{\partial \vartheta}, \ \dot{J}_2 = \left\langle -\frac{\partial H}{\partial \theta_2} \right\rangle = n_0 \frac{\partial \bar{H}}{\partial \vartheta}$$
$$\dot{\vartheta} = m_0 \frac{\partial \bar{H}}{\partial J_1} - n_0 \frac{\partial \bar{H}}{\partial J_2} \tag{7}$$

It is obvious that System (7) possesses an additional integral of motion, besides the averaged Hamiltonian (6). Let us adopt that $m_0, n_0$ are positive. Then, this additional integral may be written as follows:

$$n_0 J_1 + m_0 J_2 = N^2 = \text{const} \tag{8}$$

Eq. (8) gives rise to the trigonometric change of variables:

$$J_1 = \frac{N^2 \sin^2 \gamma/2}{n_0}, J_2 = \frac{N^2 \cos^2 \gamma/2}{m_0} \tag{9}$$

Substituting (9) into (6), one obtains:

$$h(\gamma, \vartheta) = \bar{H}(\frac{N^2 \sin^2(\gamma/2)}{n_0}, \frac{N^2 \cos^2(\gamma/2)}{m_0}, \vartheta) = \text{const}, \ \vartheta \in [0, 2\pi), \ \gamma \in [0, \pi] \tag{10}$$

The conservation law (10) guarantees that the dynamical system on the sphere ($\gamma$ is the polar angle and $\vartheta$ the azimuth angle) is completely integrable. It is important that this system is revealed explicitly without writing down the equations of motion. Dynamics of this system are described by the following simple symmetric equations:

$$\dot{\gamma} = -\frac{2m_0 n_0}{N^2 \sin \gamma} \frac{\partial h}{\partial \vartheta}; \ \dot{\vartheta} = \frac{2m_0 n_0}{N^2 \sin \gamma} \frac{\partial h}{\partial \gamma} \ . \tag{11}$$

It is self-evident that the function $h(\gamma, \vartheta)$ is the first integral for System (11), but the slow angle variables $(\gamma, \vartheta)$ do not form the canonically conjugate pair.

## 2.2 Single-DOF oscillator with periodic forcing.

Let us consider the conservative single-DOF oscillator without damping under periodic forcing with certain fixed frequency $\omega$. We also suppose that this forced system is Hamiltonian; for the most popular cases of external and parametric forcing it is easy to demonstrate that it is the case. Of course, this Hamiltonian will be time-dependent and the system will not be conservative. After the canonical transformation to the AA variables similar to (3) this Hamiltonian will take the following general form:

$$H = H(I,\theta,t); \; H(t) = H(t+T), T = 2\pi/\omega \tag{12}$$

Due to the supposed periodicity of the forcing the Hamiltonian can be expanded into the Fourier series:

$$H(I,\theta,t) = \sum_{m,n} V_{m,n}(I)\exp(i(m\theta - n\omega t)), V_{m,n} = V^*_{-m,-n} \tag{13}$$

Similarly to the treatment presented in the previous Section, we suggest that there exists the slow phase variable $\vartheta = m_0\theta - n_0\omega t, m_0, n_0 \in \mathbb{Z}$. Averaging over the fast variable and substituting the averaged actions yields:

$$\bar{H}(J,\vartheta) = \sum_{p \in Z} V_{m_0 p, n_0 p}(J)\exp(ip(m_0\theta - n_0\omega t)) \tag{14}$$

The slow evolution of the action variable is described by the following equation:

$$\dot{J} = \left\langle -\frac{\partial H}{\partial \theta}\right\rangle = -m_0 \frac{\partial \bar{H}}{\partial \vartheta} \tag{15}$$

The Hamiltonian (12) is time-dependent and therefore the following relationships are valid [1, 2]:

$$\frac{dH}{dt} = \frac{\partial H}{\partial t} \Rightarrow \frac{d\bar{H}}{dt} = \frac{\partial \bar{H}}{\partial t} = -n_0\omega\frac{\partial \bar{H}}{\partial \vartheta} \; . \tag{16}$$

Combining (15) and (16), we obtain the following integral of motion in terms of the AA variables:

$$m_0\bar{H}(J,\vartheta) - n_0\omega J = \text{const} . \tag{17}$$

Note that the averaged Hamiltonian itself **does not** yield the integral of motion for the averaged system.

## 3. Relationship to complexification – averaging approach.

The formalism of the complexification-averaging approach may be briefly summarized as follows [19, 21]: a general system of equations that describes dynamics of a set of coupled (and, generically, forced and damped) oscillators with $N$ degrees of freedom can be cast in the form

$$\ddot{u}_k = F_k(u_1,...,u_N,\dot{u}_1,...,\dot{u}_N,t). \tag{18}$$

Complex variables are introduced as follows:

$$\psi_k = \dot{u}_k + i\Omega u_k. \tag{19}$$

The frequency $\Omega$ is selected with the help of various physical reasons. For instance, in quasilinear systems it is taken to be equal to the linear frequency, and in the forced systems it is usually equal to the forcing frequency. Sometimes it is left unknown (or even considered time-varying, see, e.g. [27]) and then it is computed in the course of the treatment. From Equation (19) (with constant $\Omega$, for simplicity) one can derive:

$$u_k = \frac{-i}{2\Omega}\left(\psi_k - \psi_k^*\right); \ \dot{u}_k = \frac{1}{2}\left(\psi_k - \psi_k^*\right); \ \ddot{u}_k = \dot{\psi}_k - \frac{i}{2}\left(\psi_k + \psi_k^*\right) \tag{20}$$

Substituting (20) into (18), one obtains:

$$\dot{\psi}_k = G_k(\psi_1,\psi_1^*...,\psi_N,\psi_N^*,t) \tag{21}$$

This equation is formally equivalent to (18). However, if it is possible to justify the fast-slow decomposition in a form $\psi_k = \varphi_k \exp(i\Omega t)$, where $\varphi_k$ is a slow function of time, then one can substitute this expression to (21) and average the fast variable out. As a result, one obtains the simplified slow-flow equations

$$\dot{\varphi}_k = Q_k(\varphi_1,\varphi_1^*...,\varphi_N,\varphi_N^*) \tag{22}$$

As an example, we can consider a system of coupled Duffing oscillators described by the following equations of motion:

$$\ddot{u}_k = -u_k - u_k^3 - \varepsilon(u_k - u_{3-k}), \ k=1,2$$

Change of variables (19) with $\Omega = 1$ and subsequent averaging yields:

$$\dot{\varphi}_k = \frac{3i}{8}|\varphi_k|^2 \varphi_k + \frac{i\varepsilon}{2}(\varphi_k - \varphi_{3-k}), \ k = 1, 2 \tag{23}$$

System (23) possesses an additional integral of motion $|\varphi_1|^2 + |\varphi_2|^2 = P^2 = \text{const}$. Further change of variables $\varphi_1 = P\sin(\gamma/2)\exp(i\delta_1), \varphi_2 = P\cos(\gamma/2)\exp(i\delta_2)$ leads to the following system of equations:

$$\dot{\gamma} = -\varepsilon \sin \gamma = -\frac{2}{P^2 \sin \gamma}\frac{\partial h_1}{\partial \vartheta}; \ \dot{\vartheta} = -\frac{3}{8}P^2 \cos \gamma - \varepsilon \cot \gamma \cos \vartheta = \frac{2}{P^2 \sin \gamma}\frac{\partial h_1}{\partial \gamma}$$

$$h_1 = -\left(\frac{3}{32}P^4 \sin^2 \gamma + \frac{\varepsilon}{2}P^2 \sin \gamma \cos \vartheta\right), \ \vartheta = \delta_1 - \delta_2 \tag{24}$$

A detailed analysis of a system equivalent to (24) is presented elsewhere [20]. For our purposes it is enough to note that this system is a particular case of System (11) for $m_0 = n_0 = 1$.

This fact has simple explanation. A linear oscillator with Hamiltonian $H = \frac{p^2}{2} + \frac{\Omega^2 q^2}{2}$ induces the following well-known transformation to the action – angle variables [1]:

$$q = \sqrt{\frac{2I}{\Omega}}\sin\theta, \ p = \sqrt{2I\Omega}\cos\theta. \tag{25}$$

By identifying $q = u_k$, $p = \dot{u}_k$ and combining (19) and (25), one obtains:

$$\psi_k = \dot{u}_k + i\Omega u_k = \sqrt{2I\Omega}\cos\theta + i\sqrt{2I\Omega}\sin\theta = \sqrt{2I\Omega}\exp(i\theta) \tag{26}$$

A comparison of (26) with the transformation to complex variables mentioned above

$$\psi_{1,2} = \varphi_{1,2}\exp(i\Omega t) = P\begin{pmatrix}\sin(\gamma/2)\exp(i\delta_1)\\ \cos(\gamma/2)\exp(i\delta_2)\end{pmatrix}\exp(i\Omega t) \tag{27}$$

reveals a similarity between the CxA and general relationships (7-8). More precisely, in the case of 1:1 resonance one obtains explicit relationships between the AA and the CxA variables:

$$J_{1,2} = \frac{P^2}{2}\begin{pmatrix}\sin^2(\gamma/2)\\ \cos^2(\gamma/2)\end{pmatrix}, \ \theta_{1,2} = \delta_{1,2} + \Omega t \tag{28}$$

These relationships clearly demonstrate that the CxA approach is in fact a particular case of the transformation to action-angle variables with subsequent averaging. One can also argue that in the particular example of the coupled Duffing oscillators mentioned above it is easier to obtain equations similar to (11) (or to (24), with insignificant rescaling) directly from the Hamiltonian and without unnecessary complex transformations.

This simplification alone would be insufficient to define the AA-based averaging as separate method for analysis of the energy transport in essentially nonlinear systems. However, relationships (7-11) demonstrate that the AA formalism is not just the reformulation - it may be *more general*, than the CxA approach. The latter employs only the AA variables induced by the linear oscillator (and thus, technically, is a variation of a harmonic balance with slowly varying amplitudes, [24]). The general AA formalism is free from this restriction and can use transformations induced by any single-DOF Hamiltonian. In the next Section we are going to demonstrate that with the help of transition to the AA variables one can explore the energy transport in model systems with extreme nonlinearity, not treatable by the CxA approach.

4. **Energy transport in coupled strongly nonlinear oscillators.**

*4.1. Coupled vibro-impact oscillators.*

Let us begin with the strongest possible nonlinearity and consider a pair of identical impact oscillators, coupled by a linear spring of stiffness ε (see Figure 1).

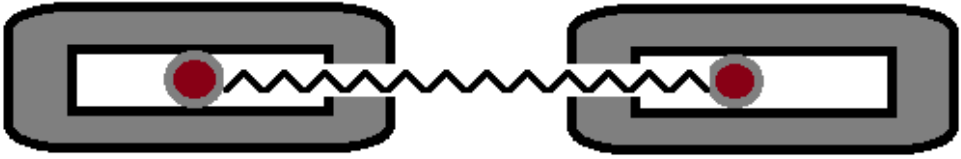

*Figure 1. Pair of impact oscillators coupled by linear spring.*

The single impactor is a particle with mass *m* moving in a channel of length 2*d*, with elastic collisions at the ends of the channel. For simplicity, it is supposed that the equilibrium length of the spring corresponds to $u_1 - u_2 = 0$. Here $u_k$ denotes the displacement of the impactor with respect to the middle point of the respective channel.

In Figure 2 we present the results of the simulation for the system depicted in Figure 1. Initially both impactors are located at the middle points of the channels, i.e. $u_1(0) = u_2(0) = 0$; initial velocity of impactor 1 is $\dot{u}_1(0) = 0.4$ (this particular value is not significant, since one can rescale the time), and the initial velocity of impactor 2 is zero, $\dot{u}_2(0) = 0$. Without restricting the generality, we suppose $m = 1, d = 1$. One can observe that for a value of coupling $\varepsilon = 0.058$ the energy remains localized at impactor 1. A minimal increase of the coupling to $\varepsilon = 0.059$ yields a qualitative change of the behavior: the impactors exchange energy. This process can be identified as nonlinear beating in the vibro-impact system.

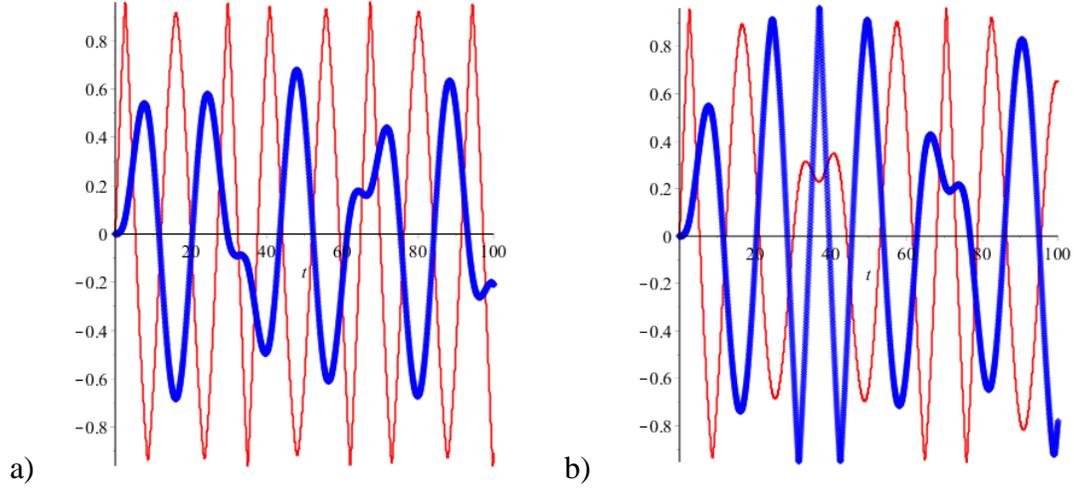

*Figure 2. Time series for displacements in coupled vibro-impact system (Figure 1) with nonzero initial velocity of one impactor; a) $\varepsilon = 0.058$, b) $\varepsilon = 0.059$. Red (thin solid) line - $u_1(t)$, blue (thick points) line - $u_2(t)$.*

To explain this transition from the localization to energy exchange, we will explore the AA formalism. Each of the impactors induces the following transformation to AA variables [4]:

$$H_k = \frac{\pi^2 I_k^2}{8md^2}, \quad u_k = \frac{2d}{\pi}\arcsin(\sin\theta_k), \quad k=1,2 \qquad (29)$$

The Hamiltonian of the system presented in Figure 1 will be expressed in terms of the AA variables as follows:

$$H = \frac{\pi^2(I_1^2 + I_2^2)}{8md^2} + \frac{2\varepsilon d^2}{\pi^2}(\arcsin(\sin\theta_1) - \arcsin(\sin\theta_2))^2 \qquad (30)$$

In order to perform the averaging, it is convenient to present the Hamiltonian (30) in the form of a Fourier series. The term $\arcsin(\sin\theta)$ represents a well-known triangular wave [28, 29]. One can easily express it as sine Fourier series and obtain

$$H = \frac{\pi^2(I_1^2 + I_2^2)}{8md^2} + \frac{32\varepsilon d^2}{\pi^4}\left(\sum_{k=0}^{\infty}\frac{(-1)^k(\sin((2k+1)\theta_1) - \sin((2k+1)\theta_2))}{(2k+1)^2}\right)^2 \qquad (31)$$

If we consider the fundamental 1:1 resonance, the slow variable will be $\vartheta = \theta_1 - \theta_2$. Averaging of the Hamiltonian (30) thus yields (up to the insignificant constant):

$$\bar{H} = \frac{\pi^2(J_1^2 + J_2^2)}{8md^2} - \frac{32\varepsilon d^2}{\pi^4}\sum_{k=0}^{\infty}\frac{\cos((2k+1)\vartheta)}{(2k+1)^4} \quad (32)$$

Transformation of the action variables according to (7-9) with $m_0 = n_0 = 1$ further yields:

$$\bar{H} = \frac{\pi^2 N^4 (\cos^4 \gamma/2 + \sin^4 \gamma/2)}{8md^2} - \frac{32\varepsilon d^2}{\pi^4}\sum_{k=0}^{\infty}\frac{\cos((2k+1)\vartheta)}{(2k+1)^4} = $$
$$= \frac{\pi^2 N^4}{8md^2}\left((1 - \frac{\sin^2 \gamma}{2}) - \kappa\sum_{k=0}^{\infty}\frac{\cos((2k+1)\vartheta)}{(2k+1)^4}\right) \quad (33)$$

The averaged Hamiltonian (33) is very simple, and it is easy to see that the structure of the phase portrait depends on a single parameter

$$\kappa = \frac{256m\varepsilon d^4}{\pi^6 N^4} \quad (34)$$

The evolution of the phase portrait on $(\vartheta, \gamma)$ surface for varying values of $\kappa$ is presented in Figure 3.

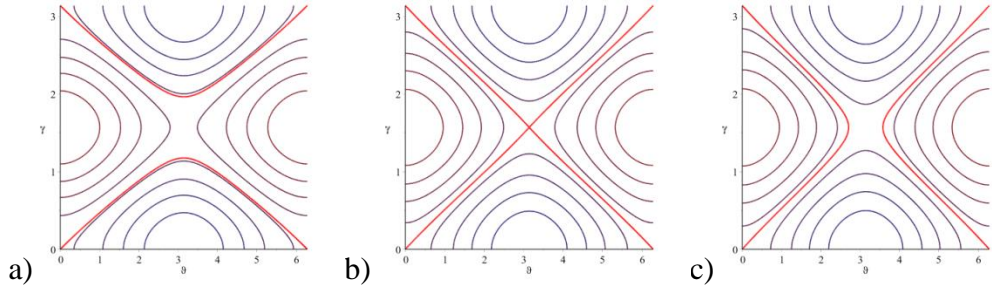

a)     b)     c)

*Figure 3. Phase portraits of the averaged system with Hamiltonian (32) for a) $\kappa = 0.22$; b) $\kappa = 24/\pi^4 = 0.2464$; c) $\kappa = 0.26$. Thick red line denotes the limiting phase trajectory (LPT).*

Initial conditions explored in the above numeric simulation (Figure 2) correspond to initial conditions $(\gamma, \vartheta) = (0, 0)$. This special orbit of averaged Hamiltonian (33) describes initial complete concentration of energy at impactor 1. To denote such orbits, L.I. Manevitch [19, 20] coined the term "limiting phase trajectory" (LPT), which will be used further on in this paper. We see that for small values of the parameter κ the LPT (thick red line in Figure 3) remains in the region $0 \leq \gamma < \pi/2$ and thus the energy is localized at impactor 1. For large values of κ the LPT covers all $0 \leq \gamma \leq \pi$ and thus energy exchange between impactors 1 and 2 (nonlinear beatings) is realized. The transition from the localization to nonlinear beatings should take place when the LPT will pass through the saddle point at $\gamma = \pi/2, \vartheta = \pi$. Values of the averaged Hamiltonian at these two points should be equal; then, one obtains the following equation for the critical value of coupling:

$$\bar{H}(0,0) = \bar{H}\left(\pi/2, \pi\right) \Rightarrow -\kappa_{cr} \sum_{k=0}^{\infty} \frac{1}{(2k+1)^4} = -\frac{1}{2} + \kappa_{cr} \sum_{k=0}^{\infty} \frac{1}{(2k+1)^4} \tag{35}$$

Taking into account the identity $\sum_{k=0}^{\infty} \frac{1}{(2k+1)^4} = \frac{15}{16}\zeta(4) = \frac{\pi^4}{96}$, one obtains:

$$\kappa_{cr} = 24/\pi^4 = 0.2464 \tag{36}$$

This value of the effective coupling corresponds to the phase portrait presented in Figure 3b. It is extremely important to note that the most interesting dynamical feature of Hamiltonian (33), i.e. the transition from localization to nonlinear beating, corresponds to passage of the LPT through the saddle point. This fact ensures slow evolution of the phase trajectories of interest in the averaged system. Thus, one can justify *a posteriori* the averaging procedure in Eqs. (30-33) despite the lack of formal small parameter. This observation is generic: closeness of the averaged phase trajectory to the saddle point can provide a slow time scale, necessary for the validity of the *ad hoc* averaging.

To compare the theoretical prediction with the numerical simulations, we first relate the value of the integral of motion *N* to the initial conditions. This parameter can be evaluated from expression for kinetic energy as follows:

$$u_1(0) = u_2(0) = 0, \dot{u}_1(0) = V_0, \dot{u}_2(0) = 0; \quad \frac{\pi^2 I_1^2(0)}{8} = \frac{\pi^2 N^4}{8} = \frac{V_0^2}{2} \Rightarrow N^4 = \frac{4V_0^2}{\pi^2} \tag{37}$$

Combining (34), (36) and (37), one obtains the following simple expression for the critical value of coupling:

$$\varepsilon_{cr} = \frac{3V_0^2}{8} \tag{38}$$

For $V_0 = 0.4$ one obtains $\varepsilon_{cr} = 0.06$ for the transition between the localization and nonlinear beatings, in excellent agreement with numeric results presented in Figure 2.

### 4.2. Coupled trigonometric oscillators.

There are few Hamiltonians that induce the transformation to AA variables in terms of elementary functions. One of them is the oscillator with Hamiltonian [4]

$$H = \frac{p^2}{2} + \frac{1}{2}\tan^2 q \tag{39}$$

Transformation of the single oscillator to the AA variables yields:

$$\begin{aligned} H &= I^2/2 + I \\ q &= \arcsin\left(\frac{\sqrt{I^2 + 2I}}{I+1}\sin\theta\right), \quad p = \frac{(1+I)\sqrt{I^2 + 2I}\cos\theta}{\sqrt{1+(I^2+2I)\cos\theta}} \end{aligned} \tag{40}$$

Let us consider the system of two such oscillators coupled through a trigonometric function, with the following Hamiltonian:

$$H = \frac{p_1^2}{2} + \frac{1}{2}\tan^2 q_1 + \frac{p_2^2}{2} + \frac{1}{2}\tan^2 q_2 + \frac{\varepsilon}{2}(\sin(q_1) - \sin(q_2))^2 \tag{41}$$

In terms of the action-angle variables, this Hamiltonian is written down as follows:

$$H = \frac{I_1^2}{2} + I_1 + \frac{I_2^2}{2} + I_2 + \frac{\varepsilon}{2}\left(\frac{\sqrt{I_1^2 + 2I_1}}{I_1 + 1}\sin\theta_1 - \frac{\sqrt{I_2^2 + 2I_2}}{I_2 + 1}\sin\theta_2\right)^2 \tag{42}$$

Considering 1:1 resonance, introducing slow variable $\vartheta = \theta_1 - \theta_2$, and performing averaging in accordance with (3-10), we obtain the following integral of motion:

$$h(\gamma,\vartheta) = -N^2 \sin^2\frac{\gamma}{2}\cos^2\frac{\gamma}{2} + \frac{\varepsilon}{4}\left( \frac{N^2\sin^4\frac{\gamma}{2}+2\sin^2\frac{\gamma}{2}}{\left(1+N^2\sin^2\frac{\gamma}{2}\right)^2} + \frac{N^2\cos^4\frac{\gamma}{2}+2\cos^2\frac{\gamma}{2}}{\left(1+N^2\cos^2\frac{\gamma}{2}\right)^2} - 2\cos\vartheta\frac{\sqrt{N^2\sin^4\frac{\gamma}{2}+2\sin^2\frac{\gamma}{2}}\sqrt{N^2\cos^4\frac{\gamma}{2}+2\cos^2\frac{\gamma}{2}}}{\left(1+N^2\sin^2\frac{\gamma}{2}\right)\left(1+N^2\cos^2\frac{\gamma}{2}\right)} \right) \quad (43)$$

Typical evolution of the phase portrait for constant $N$ and growing $\varepsilon$ is presented in Figure 4:

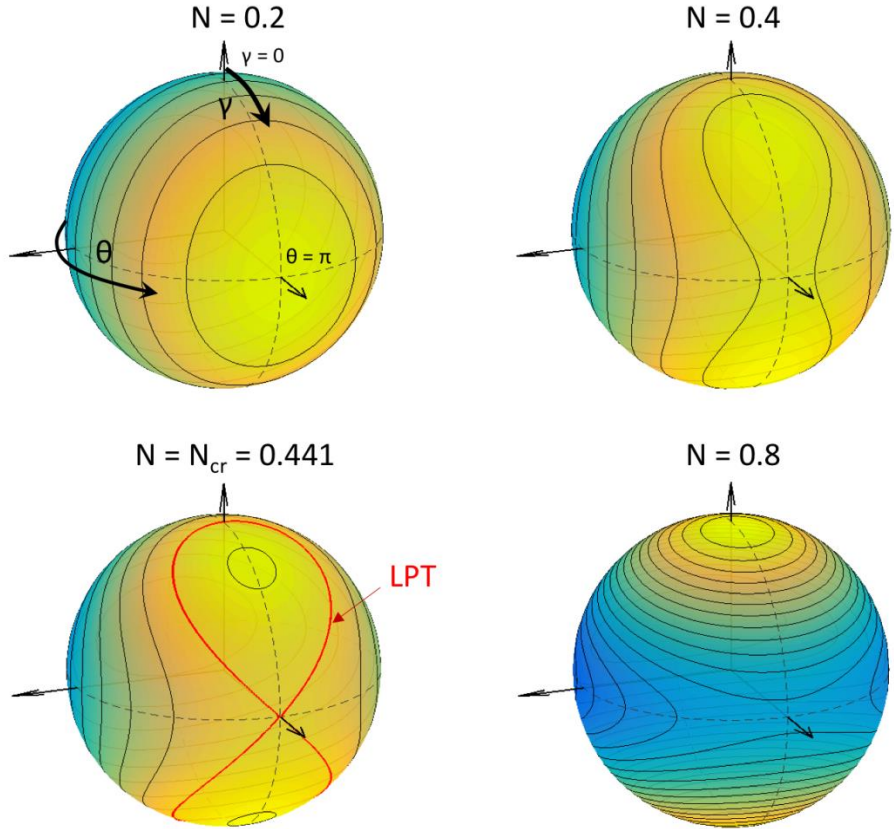

*Figure 4. Phase portrait of the effective Hamiltonian (43) on the sphere, shown in terms of color and contours in spherical coordinates, for $\varepsilon = 0.1$ and for denoted values of N.*

Here one also observes the transition from beating to localization through pitchfork bifurcation of the localized states and passage of LPT (starting from the pole $\vartheta = \pi, \gamma = 0$) through the saddle point $\vartheta = \pi, \gamma = \pi/2$. This condition translates into the following equation for critical value for transition from beatings to localization:

$$h(0,\pi) = h(\pi/2,\pi) \Rightarrow \varepsilon = \frac{N_{cr}^2 (N_{cr}^2 + 1)^2 (N_{cr}^2 + 2)^2}{3N_{cr}^6 + 18N_{cr}^4 + 24N_{cr}^2 + 8} \tag{44}$$

To check this prediction, we simulate the dynamics of the system with Hamiltonian (41). To explore the transition, we first choose $\varepsilon = 0.1$ and obtain from (44) $N_{cr} = 0.441$. Initial conditions correspond to nonzero initial displacement at the first oscillator $q_1(0) = A$ with all other IC zeros. According to (40), the value of $A$ corresponding to the transition is expressed as:

$$A_{cr} = \arcsin\left(\frac{\sqrt{N_{cr}^4 + 2N_{cr}^2}}{N_{cr}^2 + 1}\right) \tag{45}$$

Now we plot a number of time series for phase trajectories of Hamiltonian (40) for different values of $A$ (Figure 5)

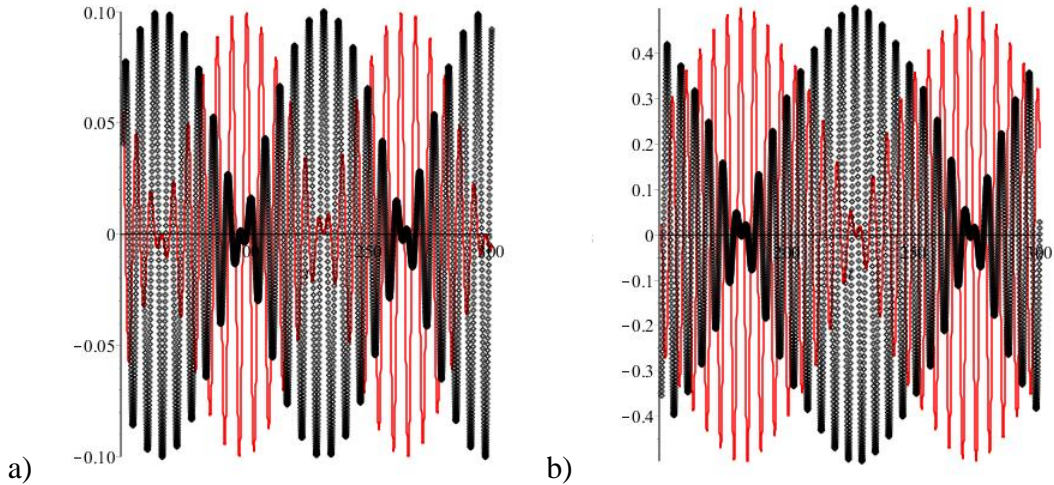

a)        b)

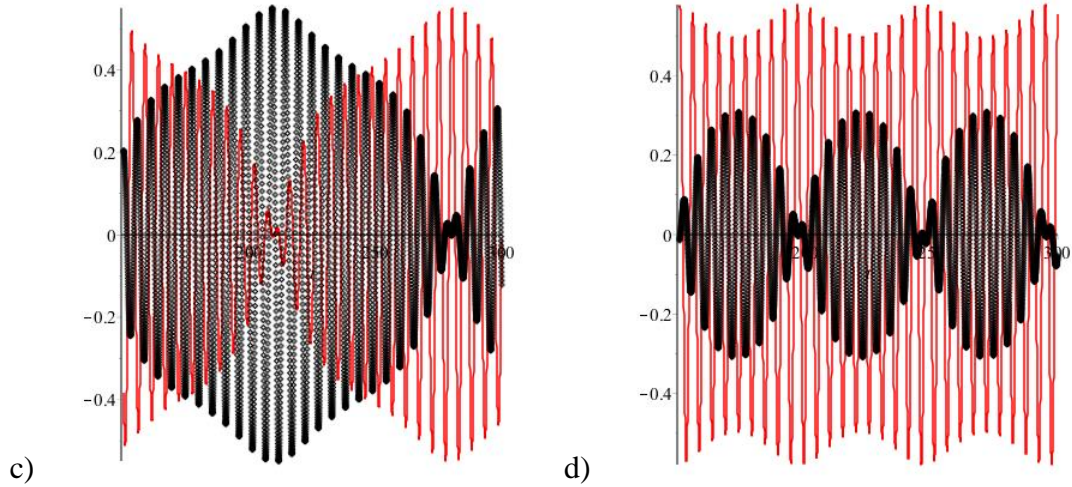

c)  d)

*Figure 5. Transition from energy exchange to localization in coupled trigonometric oscillators.. $\varepsilon = 0.1$; (a) A=0.1,( b )A=0.5, (c) A=0.55, (d) A=0.58; $q_1(t)$ - red (thin solid) line, $q_2(t)$ - black (thick point) line.*

One clearly observes the transition from the energy exchange to localization as A grows. Equation (45) predicts the transition for $A_{cr} = 0.578$, in complete agreement with numerical simulation result.

Then we explore an even larger value of the coupling parameter $\varepsilon = 2.29$ that by no means can qualify as weak coupling. The results of numeric simulation are presented in Figure 6.

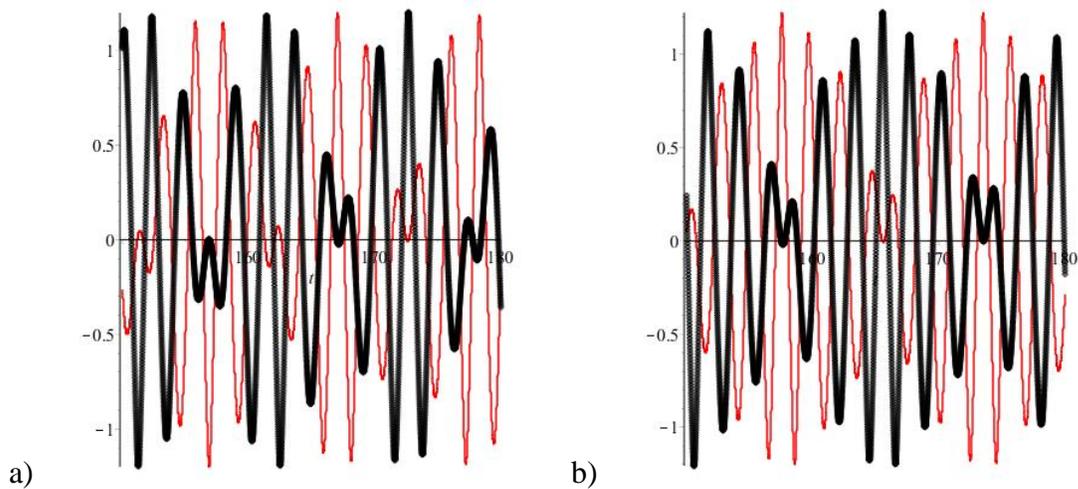

a)  b)

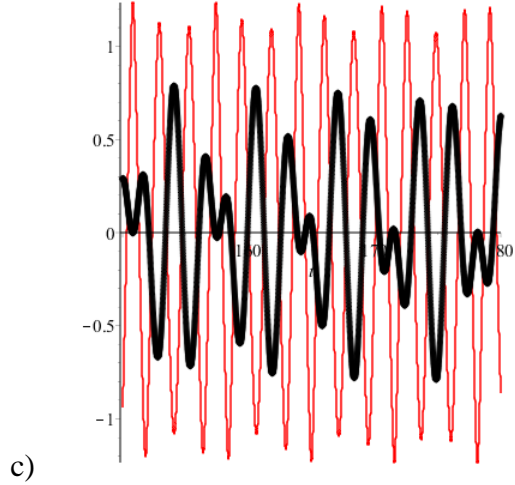

c)

*Figure 6. Transition from energy exchange to localization for extremely strong coupling, $\varepsilon = 2.29$. (a) A=1.2; (b) A=1.225; (c) A=1.235. $q_1(t)$ - red (thin solid) line, $q_2(t)$ - black (thick point) line.*

Analytic predictions (44 - 45) yield $A_{cr} = 1.258$. So, even for this large coupling the discrepancy is within 2%. This result further confirms the idea that the averaging procedure may be justified by "slow saddle dynamics" of the averaged trajectory even without the formal small parameter.

## 5. Concluding remarks

The findings presented above lead to the conclusion that the averaging based on the action-angle variables offers a convenient framework for exploration of structure and bifurcations of the slow flow, including transitions from the localization to the energy exchange. It turns out that the complexification-averaging procedure, used previously for similar problems, constitutes a particular case of a more general AA approach. In the case of the CxA, the transition to the AA variables is induced by the Hamiltonian of a linear oscillator. The AA approach is more general and allows exploration of the slow flow in systems with extreme nonlinearity, such as the coupled vibro-impact oscillators.

One can observe an interesting peculiarity of the explored systems. The averaging procedure may be justified *a posteriori*, due to slowing down the dynamics due to passage of the phase trajectory of interest close to the saddle point. Thus, the averaging procedure may be justified even in the absence of the formal small parameter. Of course, this idea has severe restrictions – such claims are valid only for the considered slow-flow phase trajectory, and not for the complete phase portrait. For instance, Figure 2 leaves the impression that the global dynamics for given set of parameters and energy level may be chaotic – like. However, even such partial information may be of considerable value, since this phase trajectory can describe important transformations in the global flow.

The authors are very grateful to the Israel Science Foundation (grant 838/13) for financial support.